\documentclass[12pt]{iopart}

\usepackage{iopams}
\usepackage {graphicx}
\begin{document}

\title{Current - voltage characteristics of break junctions of high-$T_c$ superconductors}

\author{D M Gokhfeld, D A Balaev, K A Shaykhutdinov, S I Popkov, M I Petrov}

\address{L.V. Kirensky Institute of Physics SD RAS,
Krasnoyarsk, 660036, Russia} \ead{smp@iph.krasn.ru}

\begin{abstract}

The current-voltage ($I$-$V$) characteristics of break junctions
of polycrystalline La$_{1.85}$Sr$_{0.15}$CuO$_4$,
Y$_{0.75}$Lu$_{0.25}$Ba$_2$Cu$_3$O$_{7-\delta}$,
Bi$_{1.8}$Pb$_{0.3}$Sr$_{1.9}$Ca$_2$Cu$_3$O$_x$ and composite
YBa$_2$Cu$_3$O$_{7-\delta}$ + Ag are investigated. The
experimental $I$-$V$ curves exhibit the specific peculiarities of
superconductor/normal-metal/superconductor junctions. The relation
between an $I$-$V$ characteristic of network of weak links and
$I$-$V$ dependencies of typical weak links is suggested to
describe the experimental data. The $I$-$V$ curves of typical weak
links are calculated by the K\"{u}mmel - Gunsenheimer - Nicolsky
model considering the multiple Andreev reflections.

\end{abstract}



\section{Introduction}

Measurements of current - voltage ($I$-$V$) characteristics are
accompanied with the heat emission and the selfheating. The
selfheating can modify dramatically the resulting $I$-$V$ curve. A
heat hysteresis of $I$-$V$ curve and a dependence of $I$-$V$ curve
on the velocity of scanning of current are signs of selfheating.

\begin{figure}[htbp]
\centerline{\includegraphics[width=87.5mm,height=63mm]{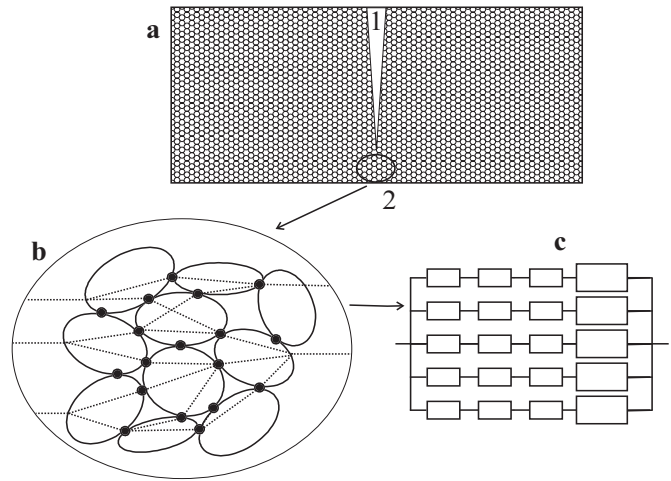}}
\caption{a) Break junction of polycrystalline sample. The crack 1
and the bottleneck 2 are displayed. b) Granules in the bottleneck.
Filled circles mark weak links that are intergranular boundaries.
Dotted lines are the main paths for transport current. c)
Simplified circuit for the network (Sec. 3.1).}
   \label{figmod}
\end{figure}

A removal of the selfheating is very important for transport
measurements of high-$T_c$ superconductors because their
heightened temperature sensibility. By reducing the cross section
$S$ of a bulk sample one can measure $I$-$V$ curve at the fixed
range of the current density for the smaller values of the
measuring current. The selfheating decreases as well. In the case
of non-tunneling break junction (BJ) technique, a significant
reducing of $S$ is achieved by the formation of a microcrack in a
bulk sample. The non-tunneling BJ of high-$T_c$ superconductors
represents two massive polycrystalline banks connected by a narrow
bottleneck (\Fref{figmod}a). The bottleneck is constituted by
granules and intergranular boundaries which are weak links
(\Fref{figmod}b). The current density in the bottleneck is much
larger than that in the banks. If the bias current $I$ is less
than the critical current $I_c$ of the bulk sample then the weak
links in the banks have zero resistance. Provided small transport
currents, (i) $I_c$ and the $I$-$V$ curve of the BJ are determined
by the weak links in the bottleneck only, (ii) the selfheating
effect is negligible.

The experimental $I$-$V$ curves of BJs of high-$T_c$
superconductors have rich peculiarities reflecting physical
mechanisms of a charge transport through weak links. It was a
topic of many investigations
\cite{zimm,svist,gonne,Pftt02,Pftt03,PphC04,Gpmm06}. Here we
analyze the earlier works of our group
\cite{Pftt02,Pftt03,PphC04,Gpmm06} and the new experimental data
(Section 2). The model for description of the $I$-$V$ curves is
suggested in Section 3. The peculiarities observed on the
experimental $I$-$V$ curves of BJs have been explained in Section
4. Also the parameters of weak links in the investigated samples
are estimated in Section 4.

\section{Experiment}

La$_{1.85}$Sr$_{0.15}$CuO$_4$ (LSCO),
Y$_{0.75}$Lu$_{0.25}$Ba$_2$Cu$_3$O$_{7-\delta}$ (YBCO) and
Bi$_{1.8}$Pb$_{0.3}$Sr$_{1.9}$Ca$_2$Cu$_3$O$_x$ (BSCCO) were
synthesized by the standard ceramic technology. The composite 67
vol.\% YBa$_2$Cu$_3$O$_{7-\delta}$ + 33 vol.\% Ag (YBCO+Ag) was
prepared from YBa$_2$Cu$_3$O$_{7-\delta}$ powder and
ultra-dispersed Ag \cite{PphC01}. The initial components were
mixed and pressed. Then the composite was synthesized at
925$^\circ$C for 8 h. The critical temperatures $T_c$ are 38 K for
LSCO, 112 K for BSCCO, 93.5 K for YBCO and YBCO+Ag.

Samples with a typical size of 2 mm x 2 mm x 10 mm were sawed out
from synthesized pellets. Then the samples were glued to a
sapphire substrates. The sapphire was chosen due to its high
thermal conductivity at low temperatures. The central part of the
samples was polished down to obtain a cross-sectional area $S
\approx 0.2$ x 1 mm$^2$. For such a value of $S$, the critical
current $I_c$ of YBCO and BSCCO has a typical value about 2 A at
4.2 K (current density $\approx 1000$ A/cm$^2$). Further
controllable decrease in $S$ is very difficult due to an
inevitable mechanical stresses breaking the sample. In order to
obtain a contact of the break junction type, the sample with the
above value of $S$ was bent together with the substrate with the
help of screws of spring-loaded current contacts. It led to the
emergence of a microcrack in the part of sample between the
potential contacts. As a result, either a tunnelling contact (no
bottleneck, the resistance $R > 10$ Ohm at the room temperature)
or a metal contact ($R < 10$ Ohm) was formed. Only the metal
contacts were selected for investigation.

The drop of $I_c$(4.2 K) when the sample was cracked shows that
the values of $S$ decreased by $\approx$30 times for LSCO and
$\approx$100 times for YBCO and BSCCO. For YBCO+Ag $I_c$(77.4 K)
decreases by $\sim$500 times.

The $I$-$V$ curves were measured by the standard four probe
technique under bias current. A typical $V(I)$ dependence of BJ
has the hysteretic peculiarity which decreases as temperature
increases. Also there is the excess current on the $I$-$V$ curves.
The $I$-$V$ curves of LSCO and BSCCO BJs exhibit an arch-like
structure at low temperature. The $I$-$V$ curves of BJs
investigated are independent of scanning velocity of bias current.
Thus, the experimental conditions provides that the hysteretic
peculiarity on these $I$-$V$ curves is not caused by the
selfheating.

\section{Model}

\subsection{$I$-$V$ curve of network}

A polycrystalline high-$T_c$ superconductor is considered to be
the network of weak links. The $I$-$V$ curve of a network is
determined by the $I$-$V$ curves of individual weak links and
their mutual disposition.

Let us consider firstly an influence of mutual disposition of weak
links on the $I$-$V$ curve. For a bulk high-$T_c$ superconductor
the $I$-$V$ curve resembles the one of typical single weak link
\cite{netw}. However the $I$-$V$ curves of BJs are distorted
usually in comparison with the one of a single weak link. It is
because the combination of finite number of weak links remains in
the bottleneck of BJ (\Fref{figmod} a, b). So the contribution of
different weak links to the resulting $I$-$V$ curve is more
stronger in a BJ than in a large network. The characteristics of a
chaotic network is difficult to calculate \cite{netw}. To simplify
the calculation of resulting $I$-$V$ curve of BJ we consider an
equivalent network: the simple parallel connection of a few chains
of series-connected weak links (\Fref{figmod} c). Indeed there are
percolation clusters \cite{perc} in a network that are paths for
current (\Fref{figmod} b). The each percolating cluster in the
considered network is considered to be the series-connected weak
links.

The $V(I)$ dependence of the series-connected weak links is
determined as $V(I) = \sum V_i(I)$, where the sum is over all weak
links in the chain, $V_i(I)$ is the $I$-$V$ curve of each weak
link. The weak links and their $I$-$V$ curves may be different. It
is conveniently to replace here the sum over all weak links with
the sum of a few more typical weak links multiplied by a weighting
coefficient $P_i$. The relation for the series-connected weak
links is resulted:

\begin{equation}
\label{eq_cha} V(I)= N_{V} \sum_i P_{i} V_{i} \left( I \right),
\end{equation}
\noindent where $N_{V}$ is the number of typical weak links, $P_i$
shows the share of $i$th weak link in the resulting $I$-$V$ curve
of the chain, $\sum P_i = 1$.

The parallel connection of chains is considered further. If the
current $I$ flows through the network then the current $I_j$
through $j$th chain equals $I P_{\parallel j} / N_\parallel$ and
$\sum I_j = I$. Here $N_\parallel$ is the number of parallel
chains in the network, $P_{\parallel j}$ is the weighting
coefficient determined by the resistance of $j$th chain, $\sum
P_{\parallel j} = 1$.

An addition (a subtraction) of chains in parallel connection
smears (draws down) the $I$-$V$ curve of network to higher (lower)
currents. It is like to the modification of $I$-$V$ curve due to
the increase (the decrease) of cross section of sample. For the
sake of simplicity the difference of parallel chains can be
neglected and the typical chain may be considered only. Then the
expression for $I$-$V$ curve of network of weak links follows:
\begin{equation}
\label{eq_netw} V(I) = N_V \sum_i P_i V_i
\left(\frac{I}{N_\parallel}\right),
\end{equation}
\noindent where the sum is over the typical weak links with
weighting coefficients $P_i$, $N_V$ is the number of
series-connected weak links in the typical chain in the network,
$I / N_\parallel = I_i$ is the current through the $i$th weak link
of the typical chain.

\subsection{$I$-$V$ curve of a typical weak link}

The metal intergranular boundaries were revealed in the
polycrystalline YBCO synthesized by the standard ceramic
technology \cite{Pftt07}. The excess current and other
peculiarities on the $I$-$V$ curves of the studied samples are
characteristic for superconductor/normal-metal/superconductor
(SNS) junctions \cite{likharev}. These facts verify that the
intergranular boundaries in the high-$T_c$ superconductors
investigated are metallic. Therefore the networks of SNS junctions
are realized in the samples.

The K\"{u}mmel - Gunsenheimer - Nicolsky (KGN) theory \cite{kgn}
only among theories developed for SNS structures predicts the
hysteretic peculiarity on the $I$-$V$ curve of weak link. The KGN
theory considers the multiple Andreev reflections of
quasiparticles. According to the KGN model, the hysteretic
peculiarity reflects a part of $I$-$V$ curve with a negative
differential resistance which can be observed under bias voltage
\cite{likharev,kgn}. The KGN approach was used earlier to the
description of experimental $I$-$V$ curves of low-$T_c$
\cite{sust07g,phC07g} and high-$T_c$ weak links
\cite{PphC99,sust07g}.

The approach based on consideration of the phase slip in nanowires
\cite{piraux} may alternatively be employed to compute the
hysteretic $I$-$V$ curve. The model \cite{piraux} is valid at $T
\approx T_c$ while the KGN model is appropriate at temperature
range $T < T_c$.

We use the simplified version \cite{sust07g} of KGN to describe
the $I$-$V$ curves of individual weak links. According to
\cite{sust07g} the expression for the current density of SNS
junction is given by:

\begin{eqnarray}
 j(V) = \sum\limits_n \exp \left( { - \frac{d}{l} n}
\right) \Biggl \lbrace \frac{e {m^\ast}^2 d^2}{2 \pi^3 \hbar^5}
\int\limits_{-\Delta + neV}^\Delta dE {\frac{\left| E \right|\sqrt
{\Delta ^2 - E^2}}{\left( 1 - C \frac{2 \left| E \right|}{\pi
\Delta} \right)^3}} \tanh{\left( \frac{E}{2k_B T} \right)}
\nonumber
\\* + \frac{e {k_F}^2 }{4 \pi^2 \hbar} \int\limits_{E_1}^{\Delta + eV}
dE \frac{E}{\sqrt {E^2 - \Delta^2}} \tanh{\left( \frac{E}{2k_B T}
\right)} \Biggr \rbrace + \frac{V}{R_{N} A} \qquad\qquad
\label{eq_ju}
\end{eqnarray}
\noindent with $C= \pi/2(1-d m^\ast\Delta/2 \hbar^2 k_{F})$ for
$C>1$ and $C=1$ otherwise, $E_{1}$ = $-\Delta + neV$ for $-\Delta
+ neV \ge \Delta $ and $E_{1}=\Delta $ otherwise. Here $A$ is the
cross section area and $d$ is the thickness of normal layer with
the inelastic mean free path $l$ and resistance $R_N$, $e$ is the
charge and $m^*$ is the effective mass of electron, $\Delta$ is
the energy gap of superconductor, $n$ is the number of Andreev
reflections which a quasiparticle with energy $E$ undergoes before
it moves out of the normal layer.

One should calculate a few $I(V)$ dependencies by Eq.(\ref{eq_ju})
for different parameters to simulate the $I$-$V$ curve of network
by Eq.(\ref{eq_netw}). Almost all parameters in Eq.(\ref{eq_ju})
can be dispersing for different weak links. Indeed there are some
distribution functions of the parameters of intergranular
boundaries ($d$, $A$, $R_N$) or the parameters of superconducting
crystallites ($\Delta$, the angle of orientation) in the SNS
network.

\section{Current - voltage characteristics}

Figures 2-5 show the experimental $I$-$V$ curves of BJs (circles)
and the calculated $I$-$V$ curves of SNS networks (solid lines).
The right scale of $V$-axis of all graphs is given in the units
$eV/\Delta$ to correlate the position of peculiarities on $I$-$V$
curve with the value of energy gap.

The parameters of superconductors are presented in Table 1. The
mean values of energy gap $\Delta_0$ at $T = 0$ known to be for
high-$T_c$ superconductors were used. Parameters $k_F, m^*$ were
estimated by the Kresin-Wolf model \cite{krw}.

\begin{table}
\label{t1} \caption{Parameters of superconductors}
\begin{indented}
\item[]
\begin{tabular}{lccccc}
\br

Sample & $\Delta_0$ [meV]& $m^*/m_e$ & $k_F$ [\AA$^{-1}$] \\

\mr

BSCCO   & 25   & 6.5 & 0.61 \\
LSCO    & 9    & 5 & 0.35 \\
YBCO    & 17.5 & 5 & 0.65 \\
YBCO+Ag & 17.5 & 5 & 0.65 \\

\br
\end{tabular}
\end{indented}
\end{table}

\begin{figure}[htbp]
\centerline{\includegraphics[width=87.5mm,height=69mm]{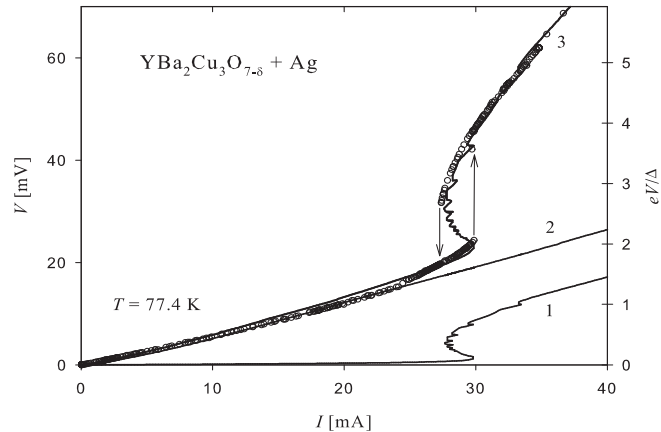}}
\caption{$I$-$V$ curve of YBCO+Ag break junction at $T$ = 77.4 K.
Experiment (circles) and computed curves (solid lines). Arrows
display the jumps of voltage drop. Curve 1 that is $N_\parallel
I_1(V_1)$ fits the hysteretic peculiarity. Curve 2 that is
$N_\parallel I_2(V_2)$ fits the initial part of $I$-$V$ curve.
Curve 3 is the dependence $V(I) = N_V \left(P_1 V_1(I/N_\parallel)
+ P_2 V_2(I/N_\parallel)\right)$.}
   \label{figAg}
\end{figure}

For a fitting we have calculated the $I(V)$ dependencies of
different SNS junctions by Eq.(\ref{eq_ju}) to describe different
parts of the experimental $I$-$V$ curve. The parameters varied
were $d$ and $R_N$. Then we have substituted the arrays of $I$-$V$
values to Eq.(\ref{eq_netw}). The most experimental $I$-$V$ curves
are satisfactory described when the sum in Eq.(\ref{eq_netw})
contains at least two members. The first member describes the
hysteretic peculiarity, the second one describes the initial part
of $I$-$V$ curve. The fitting is illustrated in detail on
\Fref{figAg} were curve 1 is calculated for $d$ = 78 \AA, curve 2
is calculated for $d$ = 400 \AA

\begin{figure}[htbp]
\centerline{\includegraphics[width=87.5mm,height=63mm]{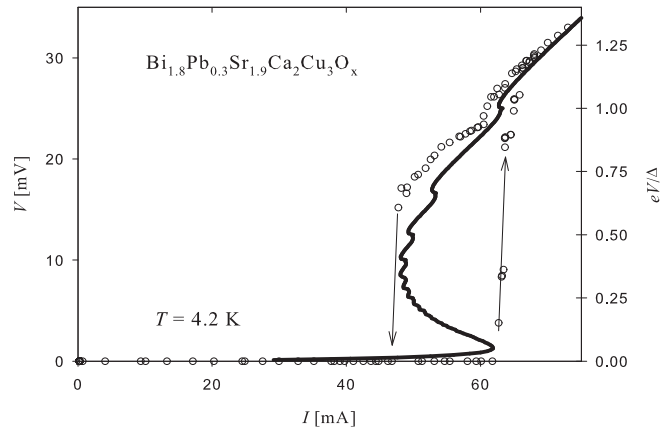}}
\caption{$I$-$V$ curve of BSCCO break junction at $T$ = 4.2 K.
Experiment (circles) and  computed curve (solid line). Arrows
display the jumps of voltage drop.}
   \label{figB}
\end{figure}

\begin{figure}[htbp]
\centerline{\includegraphics[width=87.5mm,height=63mm]{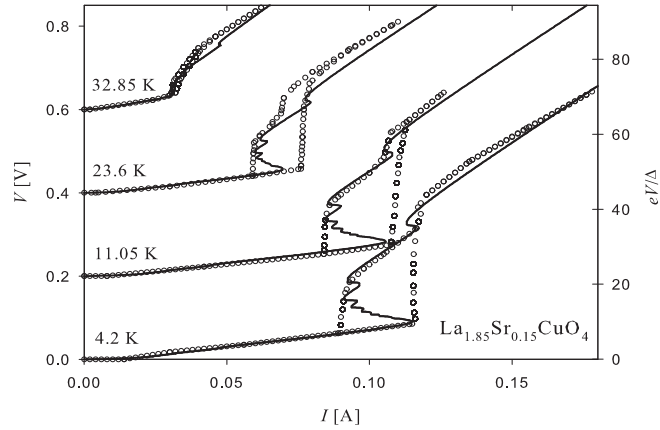}}
\caption{Temperature evolution of $I$-$V$ curve of LSCO break
junction. Experiment (circles) and computed curves (solid lines).
The $I$-$V$ curves at 11.05 K, 23.6 K, 32.85 K are shifted up by
0.2 V, 0.4 V, 0.6 V correspondingly.}
   \label{figL}
\end{figure}

\begin{figure}[htbp]
\centerline{\includegraphics[width=87.5mm,height=63mm]{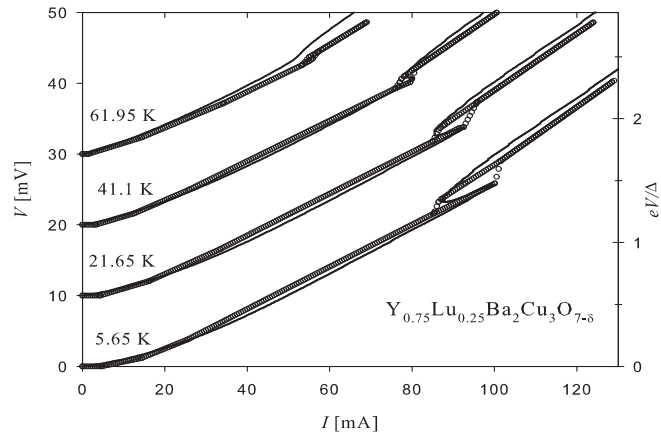}}
\caption{Temperature evolution of $I$-$V$ curve of YBCO break
junction. Experiment (circles) and  computed curves (solid lines).
The $I$-$V$ curves at 21.65 K, 41.1 K, 61.95 K are shifted up by
10 mV, 20 mV, 30 mV correspondingly.}
   \label{figY}
\end{figure}

The main fitting parameters are $d/l$, $N_V$, $N_\parallel A$,
$R_{N}/N_\parallel$, $P_{1,2}$. The parameter $P_1$ is the
weighting coefficient for the stronger (with the thinner $d$)
typical weak link, $P_2 = 1 - P_1$. Some parameters used are
presented in Table 2. Values of $l$ are estimated from the
experimental data of resistivity (2, 3, 1.6, 3.6 mOhm cm at 150 K
for bulk BSCCO, LSCO, YBCO, YBCO+Ag correspondingly) and data of
works \cite{larb,gorkop}. The value of $l$ for Ag at 77 K is known
to be $\sim 0.1$ cm. But it is more realistic to use much smaller
value for composite. Table 2 shows the possible different values
of $l$ and the corresponding values of $d$ for YBCO+Ag.

The number of the parallel paths $N_\parallel$ is estimated by
assuming $A \simeq 10^{-11}$ cm$^2$ for the weak links in
polycrystalline high-$T_c$ superconductors. Such choice of $A$ is
reasonable because the cross section area of weak link should be
more smaller than $D^2$ (\Fref{figmod}b), where $D \sim 10^{-4}$
cm is the grain size of high-$T_c$ superconductors. This rough
estimation of $N_\parallel$ is influenced by a form of the
percolation clusters in the sample \cite{perc} and imperfections
of weak links.

\begin{table}
\label{t2} \caption{Parameters of SNS junctions in the networks}
\begin{indented}
\item[]
\begin{tabular}{lccccccc} \br

Sample & $l$ [\AA] & $d_1$ [\AA] & $d_2$ [\AA]& $P_1$  & $N_V$ & $N_\parallel$ \\
\mr

BSCCO   & 72$^*$      & 3.5 & -  & 1  & 1 & 1 \\
LSCO    & 50$^*$      & 4.8 & 20 & 0.905 & 15 & 20 \\
YBCO    & 90$^*$      & 2   & 20 & 0.333 & 3 & 1 \\
YBCO+Ag & 1000$^{**}$ & 78  & 400 & 0.75 & 4 & 5 \\
        & 100$^{**}$  & 7.8 & 40 & 0.75  & 4 & 5 \\

\br
\end{tabular}
\\

$^*$ value at $T$ = 4.2 K

$^{**}$ value at $T$ = 77.4 K
\end{indented}
\end{table}

Figures 2-5 demonstrate that the hysteretic peculiarity on the
experimental $I$-$V$ curves is resulted from the region of
negative differential resistance. This region is due to the number
of the Andreev reflections decreases when the voltage increases.

The experimental $I$-$V$ curves for LSCO and YBCO at different
temperatures and the corresponding curves computed are presented
in figures \ref{figL} and \ref{figY}. We account a decreasing of
$l$ and $\Delta$ to compute $I$-$V$ curves  at higher temperatures
(for LSCO $l$(11.05 K) = 50 \AA, $\Delta$(11.05 K) = 0.93 meV,
$l$(23.6 K) = 50 \AA, $\Delta$(23.6 K) = 0.69 meV, $l$(32.85 K) =
47 \AA, $\Delta$(32.85 K) = 0.46 meV; for YBCO $l$(21.65 K) = 81
\AA, $\Delta$(21.65 K) = 17.3 meV, $l$(41.1 K) = 70 \AA,
$\Delta$(41.1 K) = 16.6 meV, $l$(61.95 K) = 60 \AA, $\Delta$(61.95
K) = 13.3 meV). The coincidence of computed curves and
experimental $I$-$V$ curves becomes less satisfactory then $T$
approaches to $T_c$. As possible, this discrepancy is due to an
influence of other thermoactivated mechanisms.

As the simulation curves demonstrate (Figs. 3 and 4), the
arch-like peculiarity on the experimental $I$-$V$ curves of LSCO
and BSCCO is one of the arches of the subharmonic gap structure
\cite{kgn}. By using Eq.(\ref{eq_netw}) we account for the
arch-like peculiarity at voltages $ \gg \Delta/e$ for LSCO
(\Fref{figL}) that should seem to contradict the KGN model
prediction for the subharmonic gap structure at $V \leq 2\Delta/e$
\cite{kgn}.

Also we have used Eq.(\ref{eq_netw}) to estimate the number of
resistive weak links in the sample of composite 92.5 vol. \% YBCO
+ 7.5 vol. \% BaPbO$_3$ \cite{PphC99}. The $I$-$V$ curve of this
composite was described earlier by the KGN based approaches
\cite{PphC99,sust07g}. We obtained $N_V$ = 13, $N_\parallel
\approx$ 4000 and the full number of resistive weak links is
52000. Small number $N_V$ is the evidence that the shot narrowest
part of bulk sample is resistive only.

\section{Conclusion}

We have measured the $I$-$V$ characteristics of break junctions of
polycrystalline high-$T_c$ superconductors. The peculiarities that
are typical for SNS junctions are revealed on the $I$-$V$ curves.

The expression for $I$-$V$ curve of network of weak links
(Eq.(\ref{eq_netw})) was suggested to describe the experimental
data. Eq.(\ref{eq_netw}) determines the relation between the
$I$-$V$ curve of network and the $I$-$V$ characteristics of
typical weak links.

The $I$-$V$ curves of SNS junctions forming the network in the
polycrystalline high-$T_c$ superconductors are described by the
K\"{u}mmel - Gunsenheimer - Nicolsky approach \cite{kgn,sust07g}.
The multiple Andreev reflections are found to be responsible for
the hysteretic and arch-like peculiarities on the $I$-$V$ curves.
The shift of subharmonic gap structure to higher voltages is
explained by the connection of a few SNS junctions in series.

We believe that the expression suggested (Eq.(\ref{eq_netw}))
allows to estimate the number of junctions with nonlinear $I$-$V$
curves and $R > 0$ in various simulated networks.

\section*{Acknowledgements}

We are thankful to R. K\"{u}mmel and Yu.S. Gokhfeld for fruitful
discussions. This work is supported by program of President of
Russian Federation for support of young scientists (grant MK
7414.2006.2), program of presidium of Russian Academy of Sciences
"Quantum macrophysics" 3.4, program of Siberian Division of
Russian Academy of Sciences 3.4, Lavrent'ev competition of young
scientist projects (project 52).

\end{document}